\documentclass[12pt,a4paper]{article}
\usepackage{latexsym,amssymb,amsmath,amsthm}

\newtheorem{theorem}{Theorem}
\newtheorem{lemma}{Lemma}

\begin{document}
\title{A relation between additive and multiplicative complexity of Boolean functions\footnote{Research supported in part by RFBR, grants
11--01--00508, 11--01--00792, and OMN RAS ``Algebraic and
combinatorial methods of mathematical cybernetics and information
systems of new generation'' program (project ``Problems of optimal
synthesis of control systems'').}}
\date{}
\author{Igor S. Sergeev\footnote{e-mail: isserg@gmail.com}}

\maketitle
\begin{abstract}
In the present note we prove an asymptotically tight relation
between additive and multiplicative complexity of Boolean
functions with respect to implementation by circuits over the
basis $\{\oplus,\wedge,1\}$.
\end{abstract}

To start, consider a problem of computation of polynomials over a
semiring $(K,+,\times)$ by circuits over the arithmetic basis
$\{+,\times\} \cup K$.

It's a common knowledge that a polynomial of $n$ variables with
nonscalar multiplicative complexity $M$ (i.e. the minimal number
of multiplications to implement the polynomial, not counting
multiplications by constants) has total complexity $O(M(M+n))$.
Generally speaking, the bound could not be improved for infinite
semirings. For instance, it follows from results by E. G.
Belaga~\cite{ebe} and V. Ya. Pan~\cite{epa} (there exist
1-variable complex and real polynomials of degree $n$ with
additive complexity $n$; at the same time, each such polynomial
has nonscalar multiplicative complexity $O(\sqrt n)$~\cite{eps}).

An analogous standard bound for finite semirings is $O(M(M+n)/\log
M)$. Generally speaking, this bound is also tight in order. A
result of such sort was proven in~\cite{ez}.\footnote{\cite{ez}
deals with monotone Boolean circuits.} We prove a similar but
asymptotically tight result.

\begin{theorem}
If a Boolean function of $n$ variables can be implemented by a
circuit over the basis $\{\oplus, \wedge, 1\}$ of multiplicative
complexity $M=\Omega(n)$, then it can be implemented by a circuit
of total complexity $(1/2+o(1))M(M+2n)/\log_2 M$ over the same
basis. The bound is asymptotically optimal.
\end{theorem}

The stated result is nearly folklore, since it's an immediate
corollary of results by E. I. Nechiporuk of early 1960s. However,
these results are little known, and the corollary is even less
known. Thus, it seems appropriate to give a proof.

The second claim of the theorem (the bound optimality) holds since
almost all Boolean functions of $n$ variables have multiplicative
complexity $\sim2^{n/2}$~\cite{en62}\footnote{Instead of this
result of Nechiporuk a trivial upper bound
$\frac3{\sqrt2}\cdot2^{n/2}$ from the later paper~\cite{ebpp} is
often cited.} and total complexity $\sim2^n/n$~\cite{elup}.

Let us prove the first claim.

Let $A$ be a Boolean matrix of size $m\times n$ ($m$ rows, $n$
columns). Assign 1 to each entry of matrix which is located at
most $\log_2 m$ positions from a one of matrix $A$ in the same
row. We denote by $S(A)$ a weight\footnote{Weight of a matrix is
the number of nonzero entries in it.} of the obtained matrix and
name it an {\it active square} of matrix $A$.

The following lemma is an appropriate reformulation of particular
case of a result due to Nechiporuk~\cite{en63,en69}. In what
follows, under an implementation of a matrix we understand an
implementation of a linear operator with that matrix.

\begin{lemma}\label{1}
Any Boolean matrix $A$ of size $m\times n$ can be implemented by
an additive circuit\footnote{Over any associative and commutative
semigroup $(G, +)$.} of complexity
$\frac{S(A)}{2\log_2m}+o\left(\frac{(m+n)^2}{\log m}\right).$
\end{lemma}

\proof Divide a set of $n$ variables into groups of $s<\log_2 m$.
All possible sums in every group can be trivially computed with
complexity $<2^s$.

Regard the computed sums as new variables and note that the
problem is now reduced to implementation of a matrix of size $m
\times 2^s\lceil n/s \rceil$ and weight $\le S(A)/s$.

Divide the new matrix into horizontal sections of height $p$.
Implement each section independently. For this, in each column of
a section group all ones into pairs. Denote by $y_{i,j}$ a sum of
(new) variables corresponding to columns with paired ones from
$i$-th and $j$-th rows.

Compute all $y_{i,j}$ independently. Next, implement an $i$-th row
of a section as $y_{i,1}+\ldots+y_{i,p}+z_i$, where $z_i$ is a sum
of variables corresponding to positions with odd ones.

Note that the total complexity of computation of all $y_{i,j}$ in
all sections is at most as large as the half of matrix weight,
that is, $S(A)/(2s)$, and the number of odd ones in each section
is at most as large as the number of columns, i.e. $2^s\lceil n/s
\rceil$. Therefore, the complexity of the described circuit is
bounded from above by
$$ \frac{n2^s}{s} + \frac{S(A)}{2s} + mp + \left\lceil \frac{m}{p} \right\rceil 2^s \left\lceil \frac{n}{s} \right\rceil. $$
Assuming $p \sim m/\log^2 m$ and $s \sim \log_2 m - 3\log_2\log_2
m$, we obtain the required bound.   \qed

The bound of lemma is asymptotically tight. More general results
of that sort established by N. Pippenger~\cite{epip} and V.~V.
Kochergin~\cite{eko}.

Now we complete the proof of the theorem. Let a circuit $S$ to
implement a Boolean function $f$ with multiplicative complexity
$M$. Number all conjunction gates in the circuit in an order not
contradicting the orientation. Denote by $h_{2i-1}, h_{2i}$ input
functions of $i$-th conjunction gate, and denote by $g_i$ its
output function.

Each function $h_j$ is a linear combination of variables and
functions $g_i$, where $1\le i < j/2$. The function $f$ itself is
a linear combination of variables and all functions $g_i$.

Computation of all functions $h_j$, $j=1,\ldots,2M$, together with
the function $f$ as linear combinations of variables and functions
$g_i$ can be performed by a linear operator with matrix of size
$(2M+1)\times(M+n)$ and active square $\le (2M+1)(n+M/2+\log_2
M)$. To obtain the desired bound, implement this operator via the
method of Lemma 1.

\end{document}


\title{О соотношении между аддитивной и мультипликативной
сложностью булевых функций\footnote{Работа выполнена при
финансовой поддержке РФФИ, проекты 11--01--00508 и 11--01--00792,
и программы фундаментальных исследований ОМН РАН <<Алгебраические
и комбинаторные методы математической кибернетики и информационные
системы нового поколения>> (проект <<Задачи оптимального синтеза
управляющих систем>>).}}
\date{}
\author{И. С. Сергеев\footnote{эл. адрес: isserg@gmail.com}}

\maketitle
\begin{abstract}
В настоящей заметке приводится доказательство асимптотически
точного соотношения между аддитивной и мультипликативной
сложностью булевых функций при реализации схемами над базисом
$\{\oplus,\wedge,1\}$.
\end{abstract}

Обратимся к задаче вычисления многочленов над некоторым
полукольцом $(K,+,\times)$ схемами над арифметическим базисом
$\{+,\times\} \cup K$.

Стандартным фактом является то, что если некоторый многочлен $n$
переменных имеет нескалярную мультипликативную сложность $M$
(минимальное число умножений в реализующей его схеме, не считая
умножений на константы), то его общая сложность не превосходит
$O(M(M+n))$. Для бесконечных полуколец эта оценка, вообще говоря,
неулучшаема, как следует, например, из результатов Э. Г.
Белаги~\cite{be} и В. Я. Пана~\cite{pa} (существуют комплексные и
вещественные многочлены одной переменной степени $n$, которые
имеют аддитивную сложность $n$; в то же время нескалярная
мультипликативная сложность любого такого многочлена не
превосходит $O(\sqrt n)$~\cite{ps}).

Для конечных полуколец стандартной является оценка $O(M(M+n)/\log
M)$, которая также, вообще говоря, неулучшаема по порядку.
Подобный результат, например, доказывается в~\cite{z}, где
рассматриваются монотонные булевы схемы. Мы приводим
асимптотически точный результат такого рода.

\begin{theorem}
Если булеву функцию $n$ переменных можно реализовать схемой над
базисом $\{\oplus, \wedge, 1\}$ мультипликативной сложности
$M=\Omega(n)$, то ее можно реализовать схемой над тем же базисом
общей сложности $(1/2+o(1))M(M+2n)/\log_2 M$. Последняя оценка
асимптотически неулучшаема.
\end{theorem}

Этот результат можно считать фольклорным, т.к. он вытекает
непосредственно из результатов Э. И. Нечипорука начала 1960-х гг.
Однако сами эти результаты, и тем более следствие из них, остаются
малоизвестными. Ввиду этого, уместно привести доказательство.

Вторая часть теоремы (оптимальность оценки) следует из того, что
почти все булевы функции $n$ переменных имеют мультипликативную
сложность $\sim2^{n/2}$~\cite{n62} (вместо этого результата
Нечипорука часто цитируется тривиальная верхняя оценка
$\frac3{\sqrt2}\cdot2^{n/2}$ из более поздней работы~\cite{bpp}) и
общую сложность $\sim2^n/n$~\cite{lup}.

Докажем первую часть.

Пусть $A$~--- булева матрица размера $m\times n$ ($m$ строк, $n$
столбцов). В каждую ячейку матрицы, расположенную не далее чем в
$\log_2 m$ позициях слева или справа от некоторой единицы матрицы
$A$, также запишем единицу.
Вес полученной матрицы обозначим $S(A)$ и назовем {\it активной
площадью} матрицы $A$ (вес матрицы --- это число ненулевых
элементов в ней).

Следующая лемма является подходящей переформулировкой частного
случая результата Нечипорука~\cite{n63,n69}. Далее, говоря о
реализации некоторой матрицы, мы имеем в виду реализацию линейного
оператора с данной матрицей.

\begin{lemma}
Булева матрица $A$ размера $m\times n$ реализуется аддитивной
схемой $($над произвольной ассоциативной и коммутативной
полугруппой $(G, +))$ сложности
$\frac{S(A)}{2\log_2m}+o\left(\frac{(m+n)^2}{\log m}\right).$
\end{lemma}

\proof Множество $n$ переменных разбивается на группы по $s<\log_2
m$ штук. Все возможные суммы в одной группе реализуются со
сложностью $<2^s$ тривиальным способом.

Если принять указанные суммы за новые переменные, то задача
сведена к вычислению матрицы размера $m \times 2^s\lceil n/s
\rceil$ и веса $\le S(A)/s$.

Эта матрица разбивается на горизонтальные полосы высоты $p$.
Матрицы в полосах реализуются независимо. Единицы в каждом столбце
такой матрицы объединяются попарно. Обозначим через $y_{i,j}$
сумму (новых) переменных, соответствующих столбцам матрицы, в
которых пару образуют единицы из $i$-й и из $j$-й строк.

Вычислим все $y_{i,j}$ независимо. Далее, любая $i$-я строка
матрицы реализуется как $y_{i,1}+\ldots+y_{i,p}+z_i$, где
$z_i$~--- сумма переменных, соответствующих позициям непарных
единиц строки.

Остается заметить, что суммарная сложность вычисления всех
$y_{i,j}$ во всех полосах не превосходит половины веса матрицы,
т.е. $S(A)/(2s)$, а число непарных слагаемых в каждой полосе не
превосходит числа столбцов $2^s\lceil n/s \rceil$. Следовательно,
сложность построенной схемы можно оценить как
$$ \frac{n2^s}{s} + \frac{S(A)}{2s} + mp + \left\lceil \frac{m}{p} \right\rceil 2^s \left\lceil \frac{n}{s} \right\rceil. $$
Полагая $p \sim m/\log^2 m$ и $s \sim \log_2 m - 3\log_2\log_2 m$,
получаем требуемую оценку.   \qed

Оценка леммы является асимптотически неулучшаемой. Более общие
результаты подобного рода получены в работах Н.
Пиппенджера~\cite{pip} и В.~В. Кочергина~\cite{ko}.

Теперь завершим доказательство теоремы. Пусть схема $S$ реализует
булеву функцию $f$ с мультипликативной сложностью $M$. Занумеруем
элементы конъюнкции в схеме в порядке, не противоречащем
ориентации. Обозначим через $h_{2i-1}, h_{2i}$ функции, подаваемые
на вход $i$-го элемента конъюнкции, а через $g_i$~--- функцию,
реализуемую на его выходе.

Любая из функций $h_j$ является линейной комбинацией переменных и
функций $g_i$, где $1\le i < j/2$. Сама функция $f$ является
линейной комбинацией переменных и всех функций $g_i$.

Вычисление всех функций $h_j$, $j=1,\ldots,2M$, и самой функции
$f$ как линейных комбинаций переменных и функций $g_i$ выполняется
линейным оператором с матрицей размера $(2M+1)\times(M+n)$,
имеющей активную площадь $\le (2M+1)(n+M/2+\log_2 M)$. Реализуя
этот оператор при помощи леммы 1, получаем оценку теоремы.